# An automatized Identity and Access Management system for IoT combining Self-Sovereign Identity and smart contracts

Montassar Naghmouchi[1][2], Hella Kaffel[1], and Maryline Laurent[2][0000-0002-7256-3721]

[1] Faculty of Science of Tunis, University of Tunis El Manar, Tunis, Tunisia
montasser.naghmouchi@etudiant-fst.utm.tn, hella.kaffel@fst.utm.tn
[2] Samovar, Télécom SudParis, Institut Polytechnique de Paris, Palaiseau, France
Maryline.Laurent@telecom-sudparis.eu

**Abstract.** Nowadays, open standards for self-sovereign identity and access management enable portable solutions that are following the requirements of IoT systems. This paper proposes a blockchain-based identity and access management system for IoT – specifically smart vehicles- as an example of use-case, showing two interoperable blockchains, Ethereum and Hyperledger Indy, and a self-sovereign identity model.

## 1  Introduction

For IoT ecosystems that require a scalable, resilient, lightweight, and secure Identity and Access Management (IAM) technologies to ensure the privacy of user-data, it is essential to implement systems that are owned directly by the device owners themselves. In the case of smart vehicles, it is logical that vehicle owners have full control over their identities, those of their vehicles and to be able to manage authorization methods since cars are private properties. Such systems are enabled with a combination of consortium or public blockchains and a Self-Sovereign Identity model (SSI).

This paper proposes a Blockchain-based IAM system that makes use of the SSI model to provide ledger-rooted identities for users and IoT devices, specifically smart vehicles. Moreover, it explores a crucial property for blockchain technology, which is interoperability. In a matter of fact, two different blockchains are used in our system - Ethereum and Hyperledger Indy - due to different capabilities provided by the two platforms.

The remaining of the paper is structured as follows. Section 2 introduces the background on SSI and related open standards. Section 3 presents our system architecture and design. Section 4 provides a full functional and security analysis and Section 5 provides an insight on blockchain interoperability and how we handle it in our proposal. Finally, we conclude our work in Section 6.

**Table 1.** Acronyms.

| Acronym | Meaning |
| --- | --- |
| DID | Decentralized Identifier |
| DPKI | Decentralized Public Key Infrastructure |
| IAM | Identity and Access Management |
| JSON-LD | JavaScript Object Notation for Linked Data |
| RWoT | Rebooting the Web of Trust |
| SSI | Self-Sovereign Identity |
| VC | Verifiable Credential |

## 2   Background on Self-Sovereign Identity and open standards

Self-Sovereign Identity (SSI) refers to the digital movement that recognizes that an individual should own and control their digital identity without relying on a third party. Online users are more aware of the value of their data and adequate privacy measures around them.

The SSI model defines the following roles:
- An ***issuer***, which is an entity that creates credentials for users.
- A ***holder***, which is a user in possession of a credential, either by ownership or by delegation from the owner.
- A ***verifier***, which is the entity that verifies a credential presented by a holder willing to obtain a service from a service provider.

The SSI model relies on four (4) key standards: Decentralized Identifiers (DID), Verifiable Credentials (VC), Decentralized Public Key Infrastructures (DPKI) and a DID Authentication protocol (DID Auth).

### 2.1   Decentralized Identifiers (DID)

DID is a W3C standard [1] that serves as an identifier for a subject. A DID is resolved into a DID Document that describes the identified subject. The DID Document is a JSON-LD data, that includes the public keys owned by the subject, service endpoints and verification methods.

A DID satisfies the following core properties: 1) permanent, i.e. which does not change or can not be re-assigned, 2) resolvable, i.e. which can be looked up to discover metadata, 3) cryptographically-verifiable, i.e. which authorship and ownership can be proved, and 4) decentralized, i.e. which do not need any central registration authority.

The DID is generated from the public key of the subject, so that the ownership of the DID can be accomplished using the private key that is cryptographically bound to a public key published in the DID Document.

### 2.2    Decentralized Public Key Infrastructure (DPKI)

Since DIDs rely on public keys, it is essential to have a Public Key Infrastructure to manage keys related to identifiers. Moreover, this infrastructure must be decentralized. DPKI or Decentralized Key Management System defines protocols to generate, store and manage public and private keys that help generate decentralized identifiers and prove ownership over them. Blockchain, as a key-value storage system can already play the role of a DPKI [2].

### 2.3    Verifiable Credentials (VC)

Verifiable credential (VC) [3] is another standard by W3C. A VC is a JSON-LD composed of assertions, about some user' identity attributes. It is issued by an issuer and held by a holder. A VC includes the issuer's public key and signature and is used to obtain services from service providers based on some claims included in that credential. It supports selective disclosure, zero-knowledge proofs and it is revocable.

### 2.4    DID Authentication protocol

DID Auth is an authentication protocol proposed by RWoT to prove the ownership or the authorship over a DID record using the authentication material specified in the DID Document (i.e. knowledge of the private key associated to the public key published in the DID document). A DID Auth process may contain verifiable credentials, as part of the exchange. DID Auth allows to establish an authenticated channel between the two parties which are usually the verifier and the holder [4].

## 3    Our system architecture and design

This section focuses on the chosen example, the blockchain-based IAM system for smart vehicles. We present the IAM model and discuss functional and security requirements. We also present the design choices and a detailed system workflow.

### 3.1    System model and vehicle sharing use case

The smart vehicle sharing use-case is considered for illustrating the need for an automated access control based on the SSI model. In this use-case, a smart vehicle owner, be it a physical person or a corporate, is able to create credentials allowing other users to gain access and usage privileges to the vehicle for various possible purposes: rental, exchange, car-sharing, vehicles for work-mission etc.

Following the SSI model, our system refers to the actors, as depicted in Figure 1:
- Vehicle owner: The owner of a vehicle is an **issuer** capable of issuing a credential for a holder authorizing access and usage of the vehicle.
- User: An entity wishing to gain access to a vehicle, once the request is made to the vehicle owner and the credential is created for the user, they are considered as a **holder** as long as the credential is valid.
- Smart contract: The **verifier** in our use-case is an Ethereum smart contract (cf. Section 3.6). The smart contract is linked to a given vehicle (one or many) and is charged of verifying the credentials presented by holders. Access is granted according to the access policies defined in the smart contract that are compared against the content of the presented verifiable credentials.
- Service Provider: The service being the vehicle usage, confers smart vehicles to service providers. These vehicles must have identifiers in the system (DIDs controlled by their owners) and have connectivity to invoke and read access decision made by verifiers, that are smart contracts.
- SSI infrastructure: Hyperledger Indy blockchain platform is used as an identity layer for our system (cf. Section 3.5). This supposes that a running Hyperledger Indy blockchain is maintained by different entities (to ensure decentralization). It can either be a consortium blockchain or a public one, depending on the implementation and real world requirements specified in a business model to ensure that vehicle owners and users are incentivized to join and use the network.
- Access Management component: Ethereum Blockchain, running smart contracts as verifiers, is the access management component and acts as an authorization layer for our system. These smart contracts can be published on the public Ethereum network and utilized by the decentralized authorization application as a blockchain back-end.

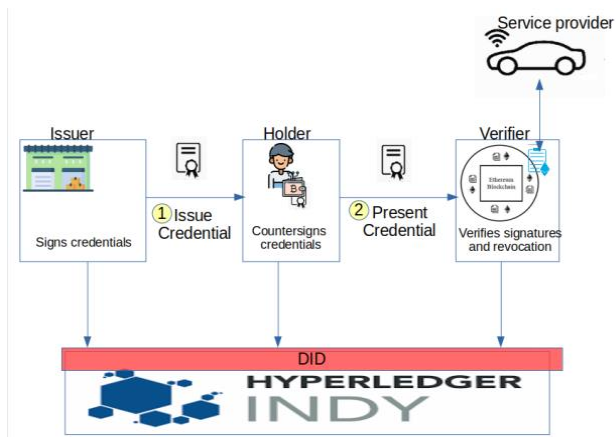

**Fig. 1.** Overview of the system architecture, along with SSI actors, other system components and standards

**3.2  Functional and security requirements**

The main requirements to be fulfilled by our system are:
- Scalability: due to the number of actors – issuers, holders, and cars – distributed over a territory and the high volume of transactions, the scalability requirement is a high priority in our system.
- Openness: the objective is to let the solution be available to as many users as possible with low entry levels, as soon as the physical persons are registered into the SSI Hyperledger Indy system.
- Availability: the IAM should remain available for serving new requests and should not be vulnerable to Single Point of Failure (SPOF) attacks.
- Automatization of the IAM: due to the number of actors, and possible transactions, this is of high interest to have the transactions processed automatically and autonomously by smart contracts.
- Accountability: Any transactions processed by our system must be logged and traced with high integrity proof guarantee for later dispute resolution.
- Flexible access policy management: access policies for vehicles should be easily updated by the owner and also fine grained for each vehicle to have a potentially customized access policy.
- Security: access to the vehicle must be conditioned by obtaining a valid credential and must not be bypassed through some fake credentials for instance.
- Privacy: identities of actors and transaction contents should remain confidential to avoid the system to leak personal data.

**3.3  Design choices**

Due to scalability, availability, traceability and automatization requirements, as identified in Section 3.2, the choice for our approach naturally fell on blockchain technologies. Moreover, the design of solution over two blockchains – Hyperledger Indy and Ethereum - and smart contract technologies were guided by the following considerations:

**Two blockchain technologies to serve the openness requirement.** The need for openness leads to the selection of public blockchains, and thus Hyperledger Indy for supporting SSI function and Ethereum for running smart contracts. Note that Hyperledger Indy is a permissioned public blockchain, while Ethereum is a public permissionless blockchain thus enabling any corporates to build any new services.

**Smart contracts to satisfy the need for IAM automatization.** Smart contracts enable to verify automatically the provided credential, for letting the smart car service know about the verification result and whether to unlock the vehicle and for writing onto Ethereum blockchain the related transaction.

### 3.4 The workflow description

All involved entities (issuers, holders, and smart vehicles) are registered in Hyperledger Indy and are provided with a DID. All the interactions between entities are setup with a DID Auth to mutually authenticate and exchange credential or credential requests as shown in Figure 2.

The issuer is responsible for publishing his smart contract(s), either one smart contract for each vehicle or one smart contract for a group of cars. The cars are configured for contacting one smart contract (at least). It might also happen that car owners cooperate to jointly publish a smart contract. Note that we can have services (in the form of smart contracts) provided by other entities should the vehicle owner chooses to use an existing smart contract developed by another party. This creates an open market for smart contract development as they can generate fees to reward developers.

The vehicles are configured for contacting at least one smart contract for authorization decisions. When a holder needs to access to the service (step 1 in Figure 2), an interaction occurs with the issuer through Hyperledger Indy, and an authenticated channel is established. The issuer can then issue a verifiable credential for the requesting holder (step 2), including specific conditions for the smart contract and the vehicle to refine the access policy for that specific holder, i.e. authorized time slots, a specific vehicle. In step 3 of Figure 2, the holder goes near to the vehicle and presents the credential contained in a mobile wallet (via NFC, Bluetooth …) which is forwarded to the Ethereum smart contract (verifier). The smart contract refers to Hyperledger Indy to verify the ownership, the authorship, and the non-revocation of the credential (step 5). The verifier also checks the credential against a list of access policies – as the credential contains claims – specified by the issuer. Note that the smart contract has exclusive invocation properties, restricted to the issuer and to the vehicles(s) controlled by that smart contract. In step 6, the smart vehicle is informed by the smart contract about the resulting decision which is written into the Ethereum blockchain. The vehicle then can unlock the doors or maintain the doors locked.

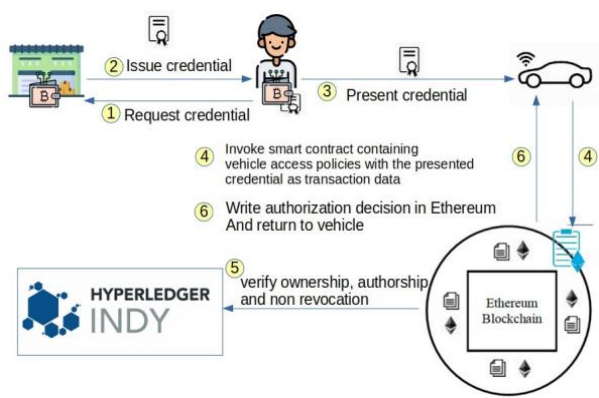

**Fig. 2.** System workflow. Describing basic steps and interactions between entities in the system.

### 3.5 Identity management with Hyperledger Indy

Hyperledger Indy is an open source project maintained by The Linux Foundation, and designed as an identity blockchain to work as an infrastructure for decentralized identities. It implements the standards introduced in Section 2 and enables a decentralized identification and authentication thanks to DIDs and public key cryptography. In the proposed system, Hyperledger Indy supports the identification function. After an identity registration transaction is received from the vehicle owner (issuer) or the vehicle renter (holder) and is validated by a trust anchor, their DID identifiers are created and registered into the blockchain. Both Issuer and Holder manage their DIDs through a wallet application which enables them to interact with the Hyperledger Indy blockchain. The wallet stores all the DIDs under the control of the wallet owner, the related private keys, the verifiable credentials, as well as the messages exchanged between entities (credential offers, credential requests, private messages).

Revocation tails for the issued credentials are also published on the ledger. These transactions represent the current state of a verifiable credential and are used to manage the revocation of credentials

### 3.6 Access management with Ethereum

The smart vehicle, considered as a service provider, delegates the credential verification and the authorization decisions to the verifier, which is an Ethereum smart contract. When a holder is physically near to the vehicle, DID Auth interaction can occur via an NFC, WiFi or Bluetooth connectivity to establish a secure communication channel between the holder and the vehicle and to enable the holder to present a verifiable presentation using claims from different credentials from their wallet. The smart car invokes the verifier smart contract on Ethereum with the verifiable presented claims as transaction data.

At this point, the smart contract verifier executes two major phases, described below in a chronological order.

**Credential Verification.** The smart contract reads data from Hyperledger Indy Blockchain to verify signatures of both issuer and holder. It also verifies the hash of the credential to ensure the integrity of its data and it verifies the non-revocation status. This processing consists of performing a lookup in Hyperledger Indy's transactions. For this purpose, there is a need for building an interactive communication method between the two blockchains to ensure interoperability (cf. Section 5).

**Accountable Authorization Decision.** The smart contract makes the authorization decision by checking the content of the verifiable credential (validity dates, allowed vehicle(s), allowed location(s), …) against the access policy specified by the vehicle owner. The authorization decision is written in Ethereum as a transaction, thus generating an access-log on Ethereum.

## 4 Functional and security analysis

**Functional analysis.** This section proposes to prove that the functional requirements identified in Section 3.2 are fulfilled by our solution:
- Flexibility: As discussed in Section 3.6, the issuer can configure an access policy in the smart contract, one for each owned vehicle or one for a group of vehicles. He can also issue a VC with an access policy customized according to the requesting holder. Both access policies can be dynamically updated, either by modifying the smart contract or the VC issuing procedure, thus ensuring a high level of flexibility.
- Scalability and availability: Issuers and holders are interacting through their wallet applications with Hyperledger Indy, whereas vehicles interact with Ethereum via smart contract invocation. This ensures a functional separation between identity related operations and authorization related operations. The fully decentralized architecture eliminates Single Point of Failure (SPOF) attacks in every step of the IAM process and provides resiliency and full time availability for actors depending on the platform. Scalability is also ensured by design, as Hyperledger Indy is a highly scalable blockchain thanks to a combination of observer nodes and validator nodes effectively handling read and write operations on the ledger. As for Ethereum, more scalability is expected with the 2.0 version of the blockchain.
- Automatized Identity and Access Management: Smart contracts present functional autonomy and a high level of automation, since they are executed when certain predefined conditions are met. This automation is seen in the authorization process as the vehicle automatically invokes the designated smart contract and the authorization decision is automatically made by the smart contract.
- Openness: Our architecture relies on two open blockchains and four open source standards. Moreover, no central authorities are consulted for the operational continuity of our solution. This lowers the entry barrier for users and eliminates any needs for third parties, making the proposal truly open.
- Accountability: Ethereum is effectively an immutable authorization decisions log, whereas Indy contains hashes for issued verifiable credentials throughout the platform history. Using the two ledgers, we can account for any authorization made on our platform with a non-disputable history.

**Security analysis.** We assume for this analysis that user wallets are not compromised for secret extraction. We also assume that there are no connectivity problems between users, vehicles and blockchains.

Authenticated channels between users are created by authenticating DIDs on Hyperledger Indy. This counteracts any attempts of masquerades or impersonation of users. Moreover, VCs are bounded to issuers and holders using signatures, this ensures that authorship and ownership can be authenticated when VCs are presented and tackles fake credentials and credential replays.

Revocation methods, like revocation tails, are also supported by VCs to ensure a secure revocation of credentials, moreover, blockchains already solve the double-spending problems of digital tokens and thus no revoked or already-used credential can be repurposed or used again.

As for identity related attacks, Hyperledger Indy is a consortium blockchain and governed by initial trust anchors called stewards that can be mainly formed of major stakeholders on the platform. Since the platform is permissioned, entry level is relatively high to ensure that the identity ledger is integer and maintained by trusted entities. At the same time, obersver nodes can be used to add transparency and accountability over these trusted entities.

As the authorization decision is made by a smart contract running on a blockchain, this decision can only be compromised by a Sybil attack targeting a majority of the Ethereum blockchain to corrupt the smart contract or alter the smart contract's decision. This makes the authorization process extremely secure as such attacks require massive resources and access and are practically impossible as they are too demanding.

Denial of service attacks are also practically impossible due to the high availability of blockchain platforms.

**Privacy analysis.** As compromised wallets are out of scope, the privacy analysis should focus on the knowledge collected by the issuer or the holder, and the leakage of information that can originate from the publicly published smart contracts, Ethereum transactions, and the published hashes of VC issued on Hyperledger Indy.

Possible traceability of user activity through published transaction should be eliminated. On another level, VC that asserts certain details about the holder must implement privacy layers since these assertions are presented to vehicles and smart contracts for authorization. Already, DID standard eliminates possible activity tracing by allowing users to own and control several DIDs at the same time. In a matter of fact, when two users interact they use relation-specific DIDs and an established secure channel via DID auth. There are as many DIDs controlled by a given user as there are relations. As for VC assertions privacy, as stated in Section 2.3, VC supports zero-knowledge proofs and selective disclosure. A user builds a verifiable presentation combining different assertions from different VCs and uses zero-knowledge to prove eligibility to verifiers, without compromising user privacy.

# 5 Blockchain Interoperability between Hyperledger Indy and Ethereum

The proposed architecture relies on two separate blockchains with separate ledgers. Interoperability in blockchain is a new trending topic in academic research since 2014 [5]. The lack of standards for blockchain systems results in Blockchain interoperability issue, which is required to ensure the openness of blockchain systems and their integration in existent systems and environments. It also permits to eliminate digital islands and contributes to enhance blockchains scalability. Interoperability resolves heterogeneity, mainly in terms of governance (public/private/consortium), openness (permissioned/permissionless), consensus algorithm and cryptographic assets like cryptocurrency and token.

## 5.1 Interoperability solutions for blockchains

With new blockchains created, each having different consensus protocols, different purposes, different capabilities and different use cases, interoperability solutions between blockchains are gaining more attention. So far, the existing methods to inter-operate blockchains are:
- Sidechains: Multiple blockchains are used to improve the system scalability. Each Multiple blockchains are used to improve the system scalability. Each Blockchain is responsible for managing a portion of the load. Generally, sidechains are secondary blockchains that are connected to a mainchain (consortium blockchain). For example, in an IoT context, Sidechains permit a finer granularity to a blockchain system where each sidechain handles a set of devices [6]. The consortium blockchain is responsible for maintaining a log of successful or failed data access requests from a consortium member to another. This can also be seen in blockchain-sharding that allows parallel transaction execution by having subsets of nodes working in parallel.
- Cross chain communication: Different blockchains are integrated in the same system or communicate with other systems, regardless of the technologies. This scenario refers to multiple needs: connect multiple blockchains to a consortium blockchain, connect blockchain engines to each other, asset swap/exchange/trading between two different blockchains, create cross-chain assets and to notify a blockchain about events happening in another blockchain. There are two cases of cross-chain asset exchange and communication: (1) Isomorphic cross-chains where the two blockchains use the same consensus algorithm, (2) Heterogeneous cross-chains where the two blockchains use different consensus algorithms [5].

Interoperability is achieved by the following components:
- Relayers: transmit messages between two chains or two blockchains. Relayers need a communication protocol to define how to transmit notifications and transactions between two blockchains. Cross-chain Communication Protocol (CCCP), is used in case of isomorphic cross-chains and Cross-Blockchain Communication Protocol (CBCP) in the case of Heterogeneous cross-chains [5].
- Notary nodes: nodes that are common between two blockchains. A notary node runs two blockchain clients at the same time. It can read from both ledgers and perform cross-chain transactions between the two blockchains. It has transaction forwarding capability. Notary nodes signs transactions with a majority of 2/3 to forward them from one blockchain A to a blockchain B. The main use case is exchanging assets.
- Other methods:  Such as smart contracts reading from another blockchain (BTC Relay) and APIs exposing transactions between two blockchains. APIs have the ability to fetch transaction data from a blockchain and expose it to be consumed by any other party, for example a web client. In our solution, we are not performing any cross-chain transactions or asset exchange between the two blockchains, so we are relying on a HTTP API to expose the Hyperledger Indy transactions to the outside world.

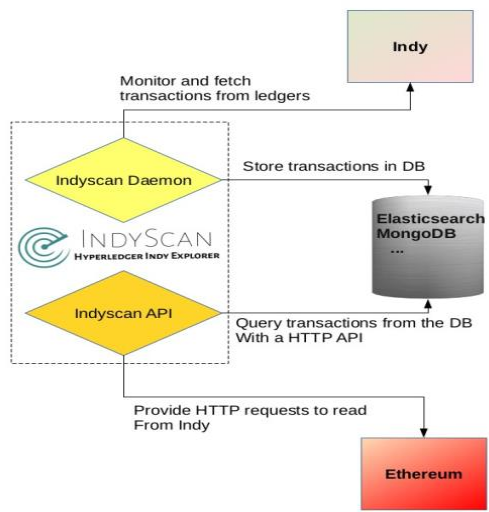

**Fig. 3.** Inter-operating Ethereum and Hyperledger Indy with APIs

### 5.2   The proposed interoperability solution

Our solution relies on a HTTP API to expose the Hyperledger Indy transactions to the outside world. Inspired by the indyscan.io [7,8], we propose the re-use of components in the repository to create a method for smart contracts on Ethereum to query transactions recorded on Hyperledger Indy. The architecture, depicted in Figure 3, relies on two important components found in the indyscan project: the indyscan daemon and the indyscan API. The indyscan daemon monitors the Hyperledger Indy blockchain and fetches transactions from the ledger. Later on, these transactions are stored in a database like MongoDB or Elasticsearch. Indyscan API then queries the database and provides a HTTP API enabling a consumer to perform HTTP requests to read from that database. Smart contracts on Ethereum consume this API, either directly or via an oracle.

## 6      Conclusion

This paper proposes an SSI compliant Identity and Access management system for a vehicle-sharing use-case. The system is based on two blockchains that are complementary, where each blockchain plays a suitable role based on its capabilities. The proposed architecture is fully decentralized and presents a system with tamper-proof identity data and permanent transparent access history for better audit and accountability. Thanks to open standards, the system is inter-operable and in compliance with modern digital identity and privacy requirements.

The choice of a self-sovereign identity model allows users to truly own and control their identities and vehicles' identities. This makes our architecture more suitable for systems with high privacy requirements. The security by design and availability features of blockchains also make this proposal a more resilient solution. The two separate blockchains with two separate ledgers provide better accountability and easier audit as blockchains have transparent data. At the same time, user's privacy is kept due to the usage of off-chain encrypted data storage, and different cryptographical methods like selective disclosure and zero-knowledge proofs. In terms of scalability, Hyperledger Indy is scalable by design with two types of nodes: validators and observers.

Managing authorization and verification with smart contracts allows for a more dynamic autonomous access control for vehicles. Access policies can be remotely updated by updating a smart contract. Some smart contracts templates can also remain publicly at the disposal of vehicle owners. The smart contracts are automated and require no intervention from humans which adds more trust between an issuer and a holder in terms of enforcing an agreement between the two parties. As for the Indy-Ethereum communication, an intermediate database is used to store the transactions from Hyperledger Indy and an HTTP API is provided for smart contracts on Ethereum to automatically verify the credentials.

Ongoing work focuses on optimizing the interoperability between the two blockchains. Furthermore, the proposal can be extended to manage identity and access for other types of devices more constrained and limited than smart cars.

**References**

1. Reed, D. et al.: Decentralized Identifiers DID 1.0 .
2. Law, J. et al.: Decentralized Key Management System Design and Architecture v4 (2019).
3. Sporny, M. et al.: Verifiable Credential Data Model 1.0 (2019).
4. Sadabello, M. et al: Introduction to DID Auth (2018).

5. Belchior, R. et al.: A Survey on Blockchain Interoperability: Past, Present, and Future Trends (2020).
6. Jiang, Y. et al.: A Cross-Chain Solution to Integrating Multiple Blockchains for IoT Data Management (2019). Sensors 2019, 19(9), 2042; https://doi.org/10.3390/s19092042
7. Indyscan website, https://indyscan.io/home/SOVRIN_MAINNET, last consulted 01/04/2021
8. Indyscan Github repository, https://github.com/Patrik-
9. Yustus Eko Oktian et al. Hierarchical Multi-Blockchain Architecture for Scalable Internet of Things. Electronics 2020, 9(6), 1050; https://doi.org/10.3390/electronics9061050